\documentstyle[graphicx,preprint,aps]{revtex}

\newcommand{\vperp}{v_{\perp}}
\newcommand{\vpar}{v_{\parallel}}
\newcommand{\pervn}{v_{\perp ,0}}
\newcommand{\parvn}{v_{\parallel, 0}}
\newcommand{\dpsi}{\Delta \psi}
\newcommand{\etal}{{\em et al}}
\newcommand{\dg}{^{\circ}}

\tightenlines

\begin{document}
\draft
\title{Enhanced Phase Space Diffusion due\\
to Chaos in Relativistic \\
Electron-Whistler Mode Wave \\
Particle Interactions with Applications  \\
to Jupiter.\\}
\author{W J Wykes \footnote{Email: wykes@astro.warwick.ac.uk \ 
Fax: +44 (0)24 76692016}, S C Chapman, G Rowlands}
\address{Space and Astrophysics Group, University of Warwick, UK}
\date{\today}
\maketitle

\begin{abstract}
\noindent

Using numerical solutions of single particle dynamics, we consider a chaotic
electron-whistler interaction mechanism for enhanced diffusion in phase space.
This process, when applied to parameters consistent with the Jovian 
magnetosphere, is a candidate mechanism for pitch angle scattering in the Io 
torus, thus providing a source of auroral precipitating electrons.

We initially consider the interaction between two oppositely directed 
monochromatic whistler mode waves. We generalize previous work to include
relativistic effects. The full relativistic Lorentz equations are solved
numerically to permit application to a more extensive parameter space. We use
this simplified case to study the underlying behaviour of the system. For
large amplitude monochromatic waves the system is stochastic, with strong
diffusion in phase space. 

We extend this treatment to consider two oppositely directed, broad band 
whistler wave packets. Using Voyager~1 data to give an estimate of the 
whistler wave amplitude at the Io torus at Jupiter, we calculate the degree 
of pitch angle scattering as a function of electron energy and initial pitch 
angle. We show that for relatively wide wave packets, significant pitch angle 
diffusion occurs (up to $\pm 25\dg$), on millisecond timescales, for electrons 
with energies from a few keV up to a few hundred keV.

\end{abstract}

\pacs{{\bf Keywords}: Relativistic, Chaos, Whistler, Pitch Angle Diffusion,
Substorms.}

\narrowtext

\section*{Introduction}

The electron-whistler interaction has been considered as potential mechanism 
for pitch angle scattering in planetary magnetospheres. Whistler waves are 
able to resonate with electrons over a broad energy range, from less than 100 
keV to several MeV \citep{horne}. In particular the Hamiltonian has been 
obtained for relativistic electrons interacting with a whistler mode wave of 
single ${\bf \hat k}$, revealing underlying behaviour that is dynamically 
simple \citep{laird}.

Gyroresonance processes with near parallel propagating whister waves have 
been considered as a pitch angle scattering mechanism (e.g. \citet{kennel}, 
\citet{lyons:84}). These processes consider electrons at resonance with many 
whistler waves of different frequencies to produce pitch angle diffusion
\citep{gendrin}. Instead we consider a process that is enhanced over (and in 
addition to) `diffusion' associated with particle dynamics strictly at 
resonance. We consider a stochastic interaction mechanism, for which
particles, that are not at resonance with the waves, effectively diffuse 
faster than the particles that are in resonance. For clarity, we will refer 
to this as `off-resonance' diffusion.

Stochasticity has been introduced by coupling the bounce motion of the trapped
electrons with a single whistler \citep{faith}, while the presence of 
a second, oppositely directed whistler wave was shown from the 
non-relativistic equations of motion to introduce stochasticity into the 
system, demonstrated numerically for a wave frequency of half the 
electron gyrofrequency by \citet{matoula}. This mechanism has been shown to
exist in self-consistent simulations \citep{devine}. Diffusion processes are 
also effected by inhomogenieties in the medium (see for example \citet{helli})
though in the case considered here variations in the background field over
the short distances traveled by the electrons during the rapid interaction 
are small and the background magnetic field can be assumed to be uniform.

Wave propagation calculations by \citet{wang} have shown that lightning 
induced whistlers are guided into the Io torus by the strong local peaks in 
density. The waves tend to oscillate around the field line corresponding to 
the peak in torus density (L=5.7, \citet{wang}), in a similar way to ducted 
whistlers with low normal angles ($\leq 10\dg$) at the equator. We consider 
the interaction of two oppositely directed whistlers with the same frequency,
propagating parallel to the background field, since small differences in the 
wave frequencies and slightly oblique waves (wave normal $\leq 10\dg$) result
in small perturbations to the dynamics of the system.

In this paper we generalize the work of \cite{matoula} to consider a range 
of wave frequencies below the gyrofrequency and include relativistic effects. 
We consider the efficiency of the mechanism in scattering electrons in 
`pancake distributions' i.e. distributions with a large peak in the 
distribution function at $90\dg$ pitch angles (sometimes represented as 
$T_{\perp} \gg T_{\parallel}$). Since we are considering single particles 
dynamic over a range of phase space, and not the evolution of the distribution
function, per se, we study pancake distributions by considering electrons 
with high initial pitch angles i.e. $\vperp > \vpar$. These then
evolve to explore phase space. Recent plasma density models have shown that 
anisotropic distributions are required to fit the observed whistler 
dispersions in the Jovian magnetosphere \citep{crary}. We investigate the 
dependence of the degree of stochasticity of the system (using Lyapunov 
exponents) on the wave amplitude, wave frequency and perpendicular velocity. 

We simplify the equations of motion in the limit of low wave amplitudes and 
extend the process to study two oppositely directed, broad band whistler wave 
packets. We obtain an estimate for the whistler wave amplitude using the 
analysis of Voyager~1 data (see \citet{scarf}, \citet{kurth} and 
\citet{hobara})) and estimate the degree of pitch angle scattering for 
different electron energies resulting from the interaction with wave packets 
of different bandwidths.

\section*{Equations of Motion}

We consider a total magnetic field of the form
\
\begin{eqnarray*}
{\bf B} = {\bf B_0} + {\bf B_{\omega}^{+}} + {\bf B_{\omega}^{-}}
\end{eqnarray*}
\
where ${\bf B_0}=B_0 {\bf \hat x}$ is the background magnetic field and
${\bf B_{\omega}^{+}}$ and ${\bf B_{\omega}^{-}}$ are the whistler waves 
propagating parallel and anti-parallel to the background field respectively 
(for coordinate system see Figure \ref{fig:geo}). We assume that the 
background field lines are uniform, since, as we will see, the interaction is 
sufficiently fast so that changes in the background field experienced by the 
electrons are small, e.g., for electrons close to Jupiter's magnetic equator 
at $6 R_J$, the field changes by less than $1\%$ for an MeV electron 
traveling at $0.9c$ and interacting with the field for 1000 electron 
gyroperiods (0.1s).

The wavefields ${\bf B_{\omega}^+}$ and ${\bf B_{\omega}^-}$ are given by
\
\begin{eqnarray*}
{\bf B}_{\omega}^+ \!\!\!&=& \!\!\!B_{\omega}
[\cos(kx-\omega t){\bf \hat y}-\sin(kx-\omega t){\bf \hat z}] \\
{\bf B}_{\omega}^- \!\!\!&=& \!\!\!B_{\omega}
[\cos(-kx-\omega t +\theta_0) {\bf \hat y} -\sin(-kx-\omega t+\theta_0)
{\bf \hat z}]
\end{eqnarray*}
\
with ${\bf \hat x}$ parallel to the background field and ${\bf \hat y}$ and 
${\bf \hat z}$ perpendicular. The wave amplitude is given by $B_{\omega}$,
while $\theta_0=\pi$ is the initial phase difference of the waves. The wave 
frequency, $\omega$, and wave number, $k$, are given by the electron whistler 
mode dispersion relation (neglecting ion effects):
\
\begin{eqnarray}
\frac{k^2c^2}{\omega^2}=1-\frac{\omega_{pe}^2}{\omega(\omega-\Omega_e)} 
\label{eq:disp}
\end{eqnarray}
\
where $\omega_{pe}$ is the plasma oscillation frequency and $\Omega_e$ is the 
electron gyrofrequency. Electrons traveling at the correct parallel 
velocity will experience a constant field and will interact strongly with it. 
This resonance velocity, ${\bf v_r}=v_r {\bf \hat x}$ is given by the 
resonance condition:
\
\begin{eqnarray}
\omega-{\bf k} \cdot {\bf v_r} = n\Omega_e /\gamma \label{eq:res}
\end{eqnarray}
\
where $n$ is an integer, and $\gamma=\left(1-v^2/c^2\right)^{-1/2}$ is the 
relativistic factor. The corresponding electric field is obtained from 
Maxwell's relation for plane propagating waves, 
$k {\bf E_{\omega}}= \omega {\bf \hat k} \wedge {\bf B_{\omega}}$ and the 
dispersion relation~(\ref{eq:disp}).

We write ${\bf v}=\vpar \,{\bf \hat x}+\vperp \cos \phi \,{\bf \hat y} +
\vperp \sin \phi \,{\bf \hat z}$, where $\phi=\phi(t)$ is the phase of the 
perpendicular velocity, define the phase angle $\psi=kx-\omega t+\phi$ 
as the angle between the perpendicular velocity and ${\bf B_{\omega}^+}$ 
and define the phase difference $\dpsi=\theta_0$ as the angle between 
the two waves.

We substitute these into the Lorentz force law to give the full equations of 
motion:
\
\begin{eqnarray}
\frac{d \vpar}{dt}&=&\frac{b \vperp}{\gamma}\left(1-\frac{\omega \vpar}{k c^2}
\right)\sin \psi +\frac{b \vperp}{\gamma}\left(1+\frac{\omega \vpar}{k c^2}
\right)\sin (\psi+\dpsi) \label{eq:f1} \\
\frac{d \vperp}{dt} &=& -\frac{b}{\gamma}\left(\vpar-\frac{\omega}{k}
\left(1+\frac{\vperp^2}{c^2}\right)\right)\sin \psi \nonumber \\
& &-\frac{b}{\gamma}\left(\vpar+\frac{\omega}{k}
\left(1+\frac{\vperp^2}{c^2}\right)\right)\sin (\psi+\dpsi) \\
\frac{d \psi}{dt} &=& k\vpar -\omega +\frac{1}{\gamma}
-\frac{b}{\gamma \vperp}\left(\vpar-\frac{\omega}{k}\right)\cos \psi 
\nonumber \\
& &-\frac{b}{\gamma \vperp}\left(\vpar+\frac{\omega}{k}\right)
\cos (\psi+\dpsi) \\
\frac{d \gamma}{dt} &=& \frac{b \omega \vperp}{k c^2}\sin \psi
-\frac{b \omega \vperp}{k c^2}\sin (\psi+\dpsi) \label{eq:f4}
\end{eqnarray}
\
where $b=B_{\omega}/B_0$ is wave amplitude scaled to the background field,
and time and velocity have been rescaled with respect to the gyrofrequency, 
$\Omega_e$, and the phase velocity, $v_{\phi}=w/k$, respectively. In the 
non-relativistic limit $\gamma \rightarrow 1$ these reduce to the 
non-relativistic equations of motion given in \citet{matoula}, 
(Equations~(5)-(7)).

\subsection*{Reduced Equations}

We can reduce the full equations of motion~(\ref{eq:f1})-(\ref{eq:f4}) to 
demonstrate the underlying properties of the system for low wave amplitudes.
The reduced equations also give qualitative information on how the system 
will be effected when different parameters are varied. To simplify the full 
equations of motion we rescale the velocities with respect to the 
normalized wave amplitude, $b$:
\
\begin{eqnarray}
\vpar &=& \parvn + b v_{\parallel ,1} \\
\vperp &=& \pervn + b v_{\perp ,1} \\
\end{eqnarray}
\
We substitute into Equations~(\ref{eq:f1})-(\ref{eq:f4}) and differentiate
with respect to time. To first order in $b$, $\parvn$ and $\pervn$ are 
constant, $\gamma_0=\sqrt{1-\parvn^2/c^2-\pervn^2/c^2}$ is a constant and
we have $\psi=kx-(1/\gamma-\omega)t$, giving the following reduced equation:
\
\begin{eqnarray}
\frac{d^2 x}{d t^2} &=& \frac{2b \pervn}{\gamma_0}
\sin\left[(1/\gamma_0-\omega)t\right]
\cos\left[kx-\theta_0\right] \label{eq:red}
\end{eqnarray}
\
Thus we have the equation of a coupled pendulum with variables 
$\vpar=\dot{x}$ and $x$. Perturbations in $\vpar$ and $x$ are proportional to 
the wave amplitude, b, to $1/\gamma_0$ and to $\pervn$.

\section*{Wave Packet Approximation}

The analysis of Voyager~1 data in \citet{scarf} and \citet{kurth} shows that
whistler mode waves at the Io torus are broad band waves. Instead of a single
pair of oppositely directed monochromatic whistler waves, we consider a sum of
N such pairs, with each pair having a different frequency and (opposite) 
wave number. For N pairs of waves, the reduced equation~(\ref{eq:red})
becomes:
\
\begin{eqnarray}
\frac{d^2 x}{d t^2} &=& \sum^N_{i=1}\frac{2\pervn}{\gamma_0}b_i
\sin\left[(1/\gamma_0-\omega_i)t\right]
\cos\left[k_i x- \theta_0,i\right]
\end{eqnarray}
\
As $N \rightarrow \infty$ we can replace the sum by an integral over 
$\omega$:
\
\begin{eqnarray}
\frac{d^2 x}{d t^2} &=& \int^{\infty}_{-\infty}\Omega_e\frac{2\pervn}
{\gamma_0}b(\omega)
\sin\left[(1/\gamma_0-\omega)t\right]
\cos\left[k(\omega)x-\theta_0(\omega)\right]d\omega
\label{eq:int}
\end{eqnarray}
\
We approximate the wave amplitude using a `top hat' distribution, i.e. we have
a wave packet with central frequency, $\omega_0$, and  constant wave amplitude 
across a narrow range of wave frequencies, $\Delta\omega$:
\
\begin{eqnarray}
b(\omega)=\left\{ \begin{array}{r@{\quad:\quad}l}
     0 & \omega<\omega_0 - \frac{\Delta\omega}{2} \\
     b_0 & \omega_0 -\frac{\Delta\omega}{2} <\omega<\omega_0 + 
\frac{\Delta\omega}{2} \\
     0 & \omega>\omega_0 + \frac{\Delta\omega}{2} 
     \end{array} \right. \label{eq:that}
\end{eqnarray}
\
We approximate the dispersion relation as 
$k(\omega)=k_0 +\beta (\omega-\omega_0)$, close to the central wave frequency,
$\omega_0$ and wave number, $k_0$, where
$\beta=\frac{3\omega_0^2-2\omega_0-(\omega_{pe}^2+k_0^2c^2)}
{2k_0c^2(\omega_0-1)}$, where $1/\beta$ is the group velocity of the wave 
packet. Substituting in and integrating with respect to the the wave 
frequency, $\omega$, gives the equation of motion for an electron interacting 
with two oppositely directed wave packets:
\
\begin{eqnarray}
\frac{d^2 x}{d t^2} &=& + \Omega_e \frac{2 b_0 \pervn}{\gamma_0}
\sin \left[(\frac{1}{\gamma_0}-\omega)t + k_0 x \right]
\frac{\sin \left[(t-\beta x)\Delta\omega/2\right]}
{(t-\beta x)} \nonumber \\
& & +\Omega_e \frac{2 b_0\pervn}{\gamma_0}
\sin \left[(\frac{1}{\gamma_0}-\omega)t - k_0 x \right]
\frac{\sin \left[(t+\beta x) \Delta\omega/2\right]}
{(t+\beta x)} \label{eq:packet}
\end{eqnarray}
\
In the limit of narrow bandwidths, $\Delta\omega\rightarrow 0$, the wave 
packet equation~(\ref{eq:packet}) yields the reduced equation~(\ref{eq:red})
with wave amplitude $b^{\prime}=b_0\Omega_e\Delta\omega$. A more detailed 
derivation of the reduced equation~(\ref{eq:red}), and the wave packet 
equation~(\ref{eq:packet}), can be found in \citet{wykes}.

\section*{Numerical Results}

Figure \ref{fig:phase} shows numerical solutions of the full 
equations of motion. The phase diagrams are stroboscopic surfaces of section 
\cite{tabor} constructed from cut-planes where $x=(n+1/2)\pi/k$, to 
sample the full electron phase space. 
The initial parallel velocity, $\vpar$, was varied over the range 
$[-v_r,v_r]$, where $v_r$ is the resonance velocity, given by the 
resonance condition (\ref{eq:res}), to give a good coverage of phase space.

All electrons were given the same initial perpendicular velocity, 
$\vperp =0.7c$ ($\vperp/v_r \approx 20$, to give high initial pitch angles 
in order to study a pancake distribution), and the phase angle, $\psi$ 
(defined as the angle between the perpendicular velocity and the first 
whistler wave ${\bf B_{\omega}^+}$, see Figure \ref{fig:geo}) was initially 
set to 0 or $\pi$.

For low wave amplitudes, Figure \ref{fig:phase}~a), the trajectories are 
essentially regular and characterized by two sets of resonances. Regular 
trajectories are described by KAM surfaces \citep{tabor} i.e. the trajectories
are near-integrable and there is an approximate constant of the motion 
associated with each orbit. The regular trajectories are represented in the 
phase diagrams as smooth lines.

As the wave amplitude is increased in Figures \ref{fig:phase}~b) and 
\ref{fig:phase}~c), stochastic effects are introduced into the region between 
the two resonances (stochastic trajectories appear as a spread of dots). For 
higher wave amplitudes, Figure \ref{fig:phase}~d), the system becomes 
increasingly stochastic with regular trajectories confined to KAM surfaces 
close to the resonances. The dependence of the degree of stochasticity on the 
wave amplitude is supported by the form of the reduced 
equation~(\ref{eq:red}), where the perturbations in parallel velocity scale 
with the wave amplitude.

In Figures \ref{fig:phase} b)--d), the stochastic regions are bounded by the 
first regular, untrapped, trajectories away from resonance. In Figures
\ref{fig:phase}~c) and \ref{fig:phase}~d), there are no regular trajectories
separating the resonances and stochastic electrons can diffuse throughout the 
stochastic region, resulting in large changes in pitch angle. It can be seen 
that the stochastic electrons are able to diffuse to a greater extent in 
phase space than the electrons on regular trajectories.

In Figure \ref{fig:vperp} we show a sequence of phase diagrams for increasing 
perpendicular velocity, with constant wave amplitude, $b=0.003$, and wave 
frequency, $\omega=1/2$. The reduced equation~(\ref{eq:red}) describes 
pendulum like behaviour with oscillations in $\vpar$ proportional 
to both the wave amplitude and the perpendicular velocity, $\vperp$. The 
resonance condition~(\ref{eq:res}) shows that 
$v_r=v_r(\gamma(\vperp, \vpar),k,\omega)$, therefore (for $\vpar << \vperp$) 
$v_r$, in addition to the total electron energy, $E$, can be described as a
function of the perpendicular velocity when $\omega$ and $k$ are constant. By 
varying $\vperp$, only we can consider the dependence of the degree of 
stochasticity on $\vperp$ and hence $E$.

In Figures \ref{fig:vperp} a) and c,) the perpendicular velocity increases 
to relativistic velocities, $\vperp$= 0.3--0.75c (Energy 25--220~keV). The 
degree of stochasticity increases with $\vperp$. From the resonance 
condition~(\ref{eq:res}) we see that increasing $\vperp$ 
increases $\gamma$ and hence reduces $v_r$ and the separation between the two 
resonances. 

In Figure \ref{fig:vperp} d) where $\vperp=\sqrt{1-\omega^2}=0.86c$ 
(E=380~keV) the resonance condition~(\ref{eq:res}) is satisfied for $v_r=0$.
Increasing $\vperp$ further causes the resonances to pass through the 
$\vpar=0$ line and change sign. In Figure~\ref{fig:vperp} e) we have $
\vperp=0.95c$ (E=740~keV). The resonance velocity now increases with 
$\vperp$ and the degree of stochasticity decreases, until the system is no 
longer dominated by stochastic trajectories (Figure~\ref{fig:vperp}~f) with 
$\vperp=0.99c$, E=1.8~MeV).

The presence of a peak in the degree of stochasticity and its dependence
on $\vperp$ is a relativistic effect. For non-relativistic velocities, 
$\gamma$, and hence the resonance velocity, is constant and the degree of 
stochasticity continually increases with $\vperp$ \citep{matoula}.

In Figures~\ref{fig:ppsi} and~\ref{fig:ptime} we show phase diagrams similar to
Figure~\ref{fig:phase} except that we now plot pitch angle against phase 
angle. The initial conditions are; $b=0.003$, $\omega=1/2$, $\vperp=0.7c$ 
and E=175~keV. The phase diagrams are qualitatively similar to 
Figure~\ref{fig:phase} and share many of the same features. Regular 
trajectories are confined to close to the resonance pitch angle,
$\alpha_r=\arctan{(\pervn/v_r)}$, where $\pervn$ is the initial 
perpendicular velocity ($0.7c$) and $v_r$ is the resonance velocity. The 
stochastic region is bounded by the first regular, untrapped, trajectories 
away from resonance. Again, we see that stochastic electrons are able to
diffuse in pitch angle to a greater extent than resonant electrons. Later, 
we shall investigate the change in pitch angle for wave amplitudes obtained
for the Jovian magnetosphere, as a function of electron energy, initial pitch
angle and wave frequency.

In Figure~\ref{fig:ptime} diffusion in pitch angle is very fast. Significant
pitch angle diffusion occurs on timescales of the order of tens of electron 
gyroperiods. On this timescale, electrons at Jupiter's magnetic equator (L=6),
traveling parallel to the background magnetic field at high relativistic 
velocities ($v\approx c$), experience changes in the background magnetic 
field of less then $1~\%$, therefore the approximation that the background 
magnetic field is uniform is valid.

In Figures~\ref{fig:wp}~a) and \ref{fig:wp}~b) we show sample numerical 
solutions of the wave packet equation~(\ref{eq:packet}) for narrow wave 
packets ($\Delta\omega=\Omega_e/500$). In panel~a) there is little change in 
pitch angle with time and in panel~b) the trajectory is qualitatively similar 
to regular trajectories in Figure~\ref{fig:ppsi}. In Figure~\ref{fig:wp}~c),
for wider wave packets, ($\Delta\omega=\Omega_e/50$), the change in pitch 
angle is greater ($\Delta\alpha\approx 15\dg$, $\alpha_0=90\dg$), with 
perturbations in pitch angle decreasing with time. The trajectory in panel~d)
is similar to stochastic trajectories in Figure~\ref{fig:ppsi}.

We have investigated solutions of the full equations of motion in order to
understand the underlying behaviour of the system. We now consider numerical
solutions of the more realistic wave packet equation for applications to the
Jovian magnetosphere.

\section*{Estimation of whistler wave amplitudes.}

We use the analysis of Voyager 1 data in \citet{scarf}, \citet{kurth} and
\citet{hobara} to estimate the whistler wave amplitude. For the Io torus at 
Jupiter the electron gyrofrequency, $\Omega_e=53.2$~kHz, corresponding to a 
background field, $B_0=302$~nT. At L=6 the plasma frequency, 
$\omega_{pe}=355$~kHz (see \citet{scarf} and \citet{kurth}). We consider 
broad band whistlers with frequencies up to the electron gyrofrequency and 
bandwidths of the order of a few kHz ($\Delta\omega=\Omega_e/50$, see 
\citet{hobara}).

The plasma wave instrument on Voyager 1 measures the electric field spectral 
density ($E_{\omega}^2/\Delta\omega$) of the whistler waves over a set of 
frequency channels of finite width $\Delta \omega$. We estimate the 
corresponding magnetic field strength using:
\
\begin{eqnarray*}
B_{\omega}=\frac{E_{\omega}}{v_{\phi}}
\end{eqnarray*}
\ 
where $v_{\phi}=\omega/k$ is the phase velocity given by the dispersion 
relation~(\ref{eq:disp}). Using a spectral density of 
$10^{-11}$~V$^2$m$^{-2}/$Hz obtained by \citet{hobara} we obtain a wave 
amplitude, $b=B_{\omega}/B_0=10^{-5}$ for a wave frequency, 
$\omega=\Omega_e/2$. While this is insufficient for stochasticity for a 
single pair of monochromatic whistlers, for broad band whistler wave packets, 
significant diffusion does occur.

We obtain a similar estimate using Ulysses data for the Jovian magnetopause
(Spectral density = $10^{-12.8}$~V$^2$m$^{-2}/$Hz for $\omega=\Omega_e/2$, see
\citet{bruce}). To determine uniquely whether or not this process is 
significant, for a given bandwidth, wave amplitude rather than spectral 
density measurements are needed.

In this context it interesting to note that for the Earth, direct amplitude 
measurements ($b=4\times 10^{-2}$, see \citet{nagano}) and the extrema of 
spectral density measurements ($b=5\times 10^{-3}$, see \citet{parrot}) yield 
whistler amplitudes sufficient for stochasticity with a pair of monochromatic 
whistlers, whereas average spectral density measurements do not 
\citep{bruce2}.

\subsection*{Pitch Angle Diffusion in the Io torus at Jupiter}

In Figure~\ref{fig:angle} we plot the change in pitch angle for electrons
during an interaction with two oppositely directed whistler wave packets with
relatively wide bandwidths ($\Delta\omega=1$~kHz, see \citet{hobara}), central
wave frequency, $\omega_0=\Omega_e/2$, and wave amplitudes consistent with the 
Jovian magnetosphere ($b=10^{-5}$). We estimate the change in pitch angle for
electron trajectories with initial pitch angles in the range 
$\alpha_0=[0\dg,180\dg]$ and electron energies from $0-400$~keV. Significant
pitch angle diffusion occurs, from a few degrees for low initial pitch 
angles, up to $25\dg$ for higher initial pitch angles, for a range of 
electron energies.

Increasing the central wave frequency has the effect of shifting the entire 
structure shown in Figure~\ref{fig:angle} down in energy. Lower wave 
frequencies are required to efficiently scatter high energy electrons 
($E>$~MeV). Given an even distribution of waves with frequencies up to the 
electron gyrofrequency, electrons with energies from a few keV up to a few 
hundreds of keV are most readily scattered. The pitch angle diffusion 
mechanism described above is consistent with the framework described by 
\citet{kennel}. However, in their derivation of the diffusion coefficient (see 
their equations $(3.6)-(3.10)$) they make an estimate of the timescale for 
diffusion based on the dynamics of particles near resonance. Here, we examine 
the detailed dynamics in phase space and find that particles off resonance 
are chaotic and as a consequence diffuse much faster than estimated by 
\citet{kennel}. Hence we have termed this process `off-resonance' diffusion.

\subsection*{Lyapunov Exponents}

Lyapunov exponents are used to quantify the degree of stochasticity in 
the system. The Lyapunov exponents are calculated using the method described 
by \citet{bene} and \citet{parker}. All six Lyapunov exponents were 
calculated over phase space and evolved to their asymptotic limit. The only 
significant Lyapunov exponent corresponds to the coordinate directed along
the background field. 

For positive Lyapunov exponents, two trajectories that are initially 
close together in phase space will diverge exponentially in time. For negative
or zero Lyapunov exponents, two trajectories that are initially close together
in phase space will remain close together. Positive (negative) Lyapunov 
exponents correspond to stochastic (regular) trajectories in phase space.
For a detailed description of the theory of Lyapunov exponents see 
\citet{hilborn}.

In Figure~\ref{fig:lyap} we plot the Lyapunov exponents for 10~keV electrons, 
averaged over a series of trajectories with pitch angles in the range 
$[0\dg,180\dg]$. We consider the interaction between electrons and two 
oppositely directed whistler wave packets with relatively wide bandwidths
($\Delta\omega=1$~kHz, see \citet{hobara}) and investigate the dependence of 
the Lyapunov exponent on the relativistically correct wave frequency,
$\omega_{\gamma}=\gamma\omega$. The Lyapunov exponents are plotted as a 
function of $1/\omega_{\gamma}$. The Lyapunov exponent is enhanced when 
$\omega_{\gamma}=1/n$, where $n$ is an integer. It then follows that close to 
these frequencies the process will be the most efficient in pitch angle
scattering. The structure in the dependence of the Lyapunov exponent on the 
relativistically correct wave frequency is subtle and arises from higher order
terms in the reduced equation~(\ref{eq:red}), a derivation of which can be 
found in \citet{wykes}.

The separation of two electrons, with Lyapunov exponent $\lambda$, in phase 
space, scales approximately as $\sim \exp{\lambda t}$, thus an order of 
magnitude estimate shows that changes in pitch angle scale approximately as 
$\sim \exp{\lambda t}$, giving a characteristic time constant, $\tau$, for 
changes in pitch angle of by a factor of $e^1$, of approximately 
$\tau\sim1/\lambda$ gyroperiods. Thus the Lyapunov exponents in 
Figure~\ref{fig:lyap} indicate changes in pitch angle occurring in tens of
electron gyroperiods.

\section*{Discussion}

We have considered the interaction between two oppositely directed whistler 
waves and relativistic electrons. We initially considered monochromatic 
whistlers in order to understand the underlying behaviour of the system. For 
whistlers with wave amplitudes consistent with the Jovian magnetosphere the 
interaction with two oppositely directed monochromatic whistler waves results 
in weak diffusion in phase space. Stochastic effects can be introduced with
high wave amplitudes: electrons not in resonance with either of the two 
whistlers can diffuse extensively in phase space. We refer to this process as 
`off-resonance' diffusion as only the electrons not in resonance are 
scattered in this way. Resonance diffusion remains unchanged.

We have extended the treatment of monochromatic whistlers to consider two 
oppositely directed whistler wave packets. For relatively broad band waves
with amplitudes consistent with quiet times in the Jovian magnetosphere, we 
have shown that the `off-resonance' diffusion results in significant pitch 
angle diffusion (up to $25\dg$) on timescales of the order of a few tens of
electron gyroperiods (milliseconds). `Off-resonance' diffusion is most
effective in scattering electrons with energies from a few keV up to a few
hundred keV (MeV electrons require extremely low wave frequencies for 
effective scattering).

We have shown that the Lyapunov exponent is enhanced when the relativistically
correct wave frequency $\omega_{\gamma}=\gamma\omega=1/n$ where $n=1,2,3,...$
This phenomenon arises purely from the interaction of the two whistler 
waves. We expect enhanced phase space diffusion close to these frequencies.

\noindent
{\bf Acknowledgements} W J Wykes and S C Chapman  are funded by PPARC. \\

\pagestyle{empty}

\begin{figure}
\caption{\label{fig:geo} Illustration of the coordinate system used
in the model. In a) the magnetic wavefields ${\bf B_{\omega}^+}$ and
${\bf B_{\omega}^-}$ lie in the ${\bf \hat y,\hat z}$ plane, perpendicular to
the background field, ${\bf B_0}=B_0 {\bf \hat x}$.
The phase angle $\psi$ is defined as the angle between ${\bf B_{\omega}^+}$
and the electron perpendicular velocity, ${\bf \vperp}$ and $\Delta \psi$ is
the angle between ${\bf B_{\omega}^+}$ and ${\bf B_{\omega}^-}$. In b) the
electron pitch angle $\alpha$ is defined as the angle between the velocity
vector ${\bf v}$ and the background field ${\bf B_0}$.}
\end{figure}

\begin{figure}[htbp]
\caption{\label{fig:phase} Stroboscopic surface of section
plots for $\omega/\Omega_e=1/2$, initial $\vperp=0.7c$. The
parallel velocity, $v_{\parallel}$, has been scaled to the phase velocity,
$v_{phase}$. For this wave frequency, the dispersion relation
(\ref{eq:disp}), gives $v_{phase}=0.07c$. The phase angle $\psi$ is defined
as the angle between the perpendicular velocity and the whistler wave
propagating in the +ve ${\bf \hat x}$ direction.
For low wave amplitudes, panel a), all trajectories are regular and the
equations of motion reduce to a pendulum equation, with resonances given by
the resonance condition (\ref{eq:res}). For slightly higher wave amplitudes,
panels b) and c), stochastic effects appear as regular trajectories are
broken down. For high wave amplitudes, panel d), phase space is dominated by
stochastic trajectories with regular trajectories confined to KAM surfaces
close to the resonances. The stochastic region is bounded above and below by 
the first regular, untrapped, trajectories away from resonance, therefore 
there is a limit on the diffusion of electrons in phase space.}
\end{figure}

\begin{figure}[htbp]
\caption{\label{fig:vperp} Stroboscopic surface of section plots for
$\omega/\Omega_e=1/2$ and $b=B_{\omega}/B_0=0.003$. Electron energy increases
as a function $\vperp$. In panels a) - c) the degree of stochasticity 
increases with increasing perpendicular velocity (and electron energy, $E$)
with constant $\omega,\Omega_e$, and $k$. As $\vperp$ and $\gamma$ increases 
the resonance velocity, $v_r$, decreases (see the resonance condition 
(\ref{eq:res})). In panel d) the resonance velocity is zero. Here 
$\vperp=0.86c$ and $E=300$~keV. Increasing $\vperp$ (and $E$) further in 
panel e) causes the resonance velocities to cross the $\vpar=0$ line and the 
degree of stochasticity now decreases, until in panel f) with $\vperp=0.99c$ 
($E=1.8 MeV$) the trajectories become regular once more.}
\end{figure}

\begin{figure}[htbp]
\caption{\label{fig:ppsi} Pitch angle plotted against phase angle for
$b=0.003$, $\omega/\Omega_e=1/2$, $\vperp=0.7$ and $E=175$~keV. Phase space 
is divided into stochastic and regular regions in a similar way to Figure 
\ref{fig:phase}. Electrons with regular trajectories close to the velocity 
resonances in Figure \ref{fig:phase} are confined to close to the resonance 
pitch angle $\alpha_r=\arctan( {\vperp}_0/v_r)$. Pitch angle diffusion is 
limited by the extent of the stochastic region, which is bounded by the first 
untrapped orbit away from the resonances. The electrons that diffuse to the 
highest $\vpar$ in Figure \ref{fig:phase} have the greatest change in pitch 
angles.}
\end{figure}

\begin{figure}[htbp]
\caption{\label{fig:ptime} Series of phase plots showing diffusion in pitch
and phase angle for a small set of initial conditions. Diffusion is rapid
with electrons reaching the minimum pitch angle in tens of electron 
gyroperiods. After 100 electron gyroperiods, electrons have diffused 
throughout the stochastic region. Parameters are as in Figure \ref{fig:ppsi}.}
\end{figure} 

\begin{figure}[htbp]
\caption{\label{fig:wp} Changes in pitch angle for sample solutions of the 
wave packet equation for two widths of wave packet. In panels a) and b) 
$\Delta\omega = \Omega_e/5000=10Hz$ resulting in very narrow wave packets. 
Trajectories are regular and resulting changes in pitch angle are low.
In panels c) and d) $\Delta\omega = \Omega_e/50=1 kHz$, with initial pitch
angle $\alpha(t=0)=90^{\circ}$. The trajectories are now stochastic, with 
greater change in pitch angle. In panel c) it can be seen that perturbations 
in the parallel velocity decrease with time.}
\end{figure}

\begin{figure}[htbp]
\caption{\label{fig:angle} Pitch angle diffusion for relatively wide, 
oppositely directed whistler wave packets ($\Delta\omega=\Omega_e/50$) with 
central frequency $\omega=\Omega_e/2$ and low wave amplitude ($b=10^{-5}$). We 
plot the change in pitch angle against initial pitch angle and electron 
energy. Significant pitch angle diffusion occurs (from a few degrees up to 
$25^{\circ}$) for a range of electron energies and initial pitch angles.}
\end{figure}

\begin{figure}[htbp]
\caption{\label{fig:lyap} We investigate the
dependence of the Lyapunov exponent on the relativistically correct wave 
frequency $\omega_{\gamma}=\gamma\omega$, for relatively wide band whistler 
wave packets. We plot the Lyapunov exponent as a function of 
$1/\omega_{\gamma}$. The Lyapunov exponent is enhanced when 
$\omega_{\gamma}=1/n$, where $n$ is an integer.} 
\end{figure}

\begin{figure}[htbp]
\setlength{\unitlength}{0.6mm}
\begin{center}
\begin{picture}(200,150)(-100,-75)
\put(-100,-75){\framebox(200,150)}

\put(-75,-50){\begin{picture}(100,100)(-25,-50)
\put(-25,0){\vector(1,0){75}}
\put(0,-50){\vector(0,1){100}}
\put(0,0){\vector(1,2){25}}
\put(0,0){\vector(1,1){39}}
\put(0,0){\vector(1,-1){25}}
\qbezier(10,10)(20,0)(10,-10)
\put(10,10){\vector(-1,1){0}}
\put(10,-10){\vector(-1,-1){0}}
\qbezier(9,19)(12.4,17.2)(15,15)
\put(9,19){\vector(-2,1){0}}
\put(15,15){\vector(1,-1){0}}
\put(50,-10){$\bf \hat y$}
\put(-10,50){$\bf \hat z$}
\put(43,43){$\bf B_{\omega}^+$}
\put(27,53){$\bf B_{\omega}^-$}
\put(30,-30){$\bf v_{\perp}$}
\put(15,5){$\psi$}
\put(16,30){$\Delta \psi$}
\put(-35,45){$\odot$}
\put(-35,30){$\odot$}
\put(-35,15){$\otimes$}
\put(-25,45){$\bf B_0$}
\put(-25,30){$\bf +k$}
\put(-25,15){$\bf -k$}
\put(-25,-50){a)}
\end{picture}}

\put(-12.5,-50){\begin{picture}(100,100)(-50,-50)
\put(-25,0){\vector(1,0){75}}
\put(0,-50){\vector(0,1){100}}
\put(0,0){\vector(2,1){40}}
\put(0,0){\vector(1,0){40}}
\put(0,0){\vector(0,1){40}}
\qbezier(15,0)(14.6,3.45)(13.4,6.7)
\put(15,0){\vector(0,-1){0}}
\put(13.4,6.7){\vector(-1,2){0}}
\put(50,-10){$\bf \hat x$}
\put(-20,50){$\bf \hat y,\hat z$}
\put(-10,15){$v_{\perp}$}
\put(30,-20){$v_{\parallel}$}
\put(40,20){$\bf v$}
\put(17.5,2.5){$\alpha$}
\put(42.5,-20){$\bf B_0$}
\put(-25,-50){b)}
\end{picture}}

\end{picture}
\end{center}
\end{figure}

\begin{figure}[htbp]
\centering
\includegraphics[width=1.0\textwidth]{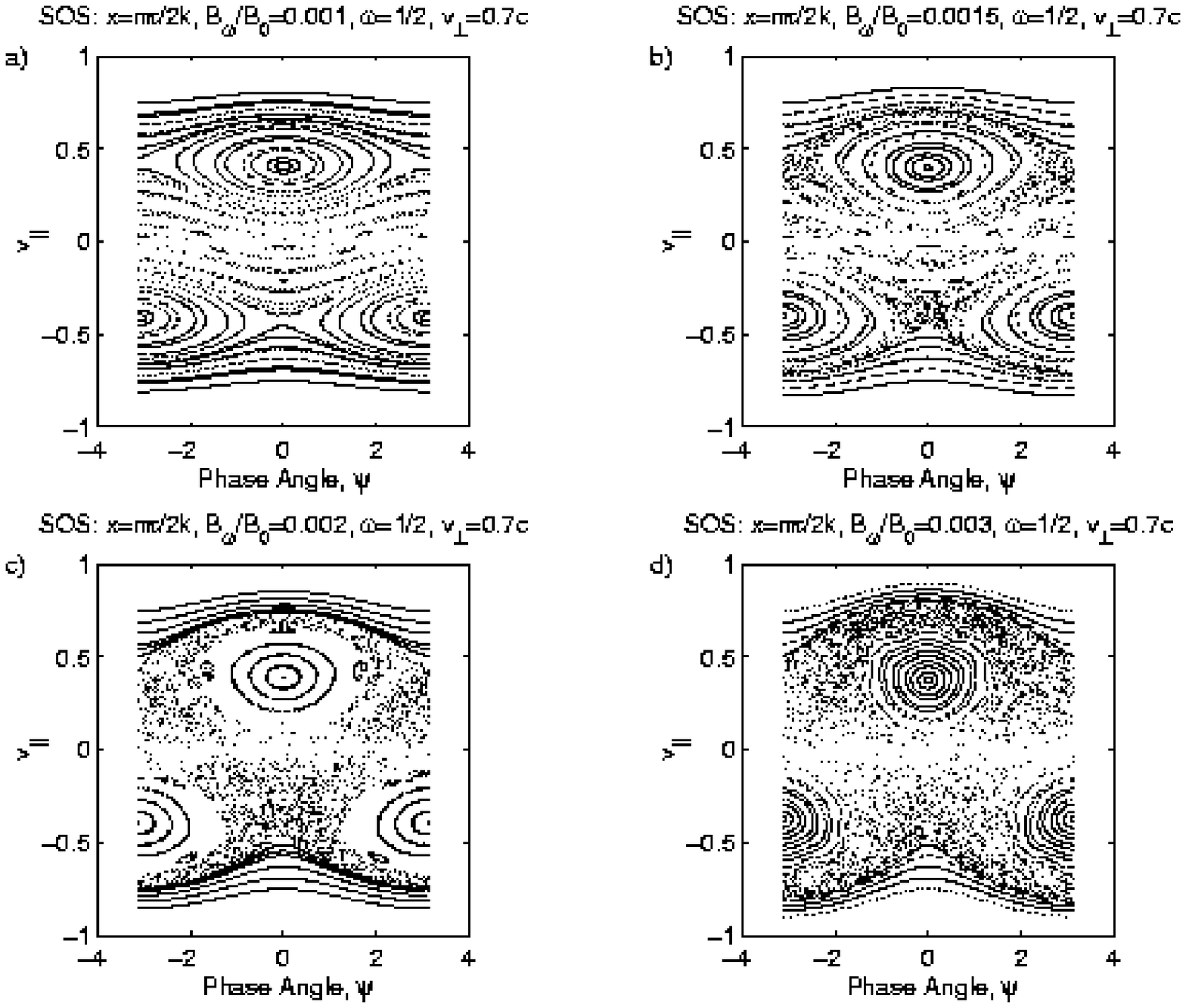}
\end{figure}

\begin{figure}[htbp]
\centering
\includegraphics[width=1.0\textwidth]{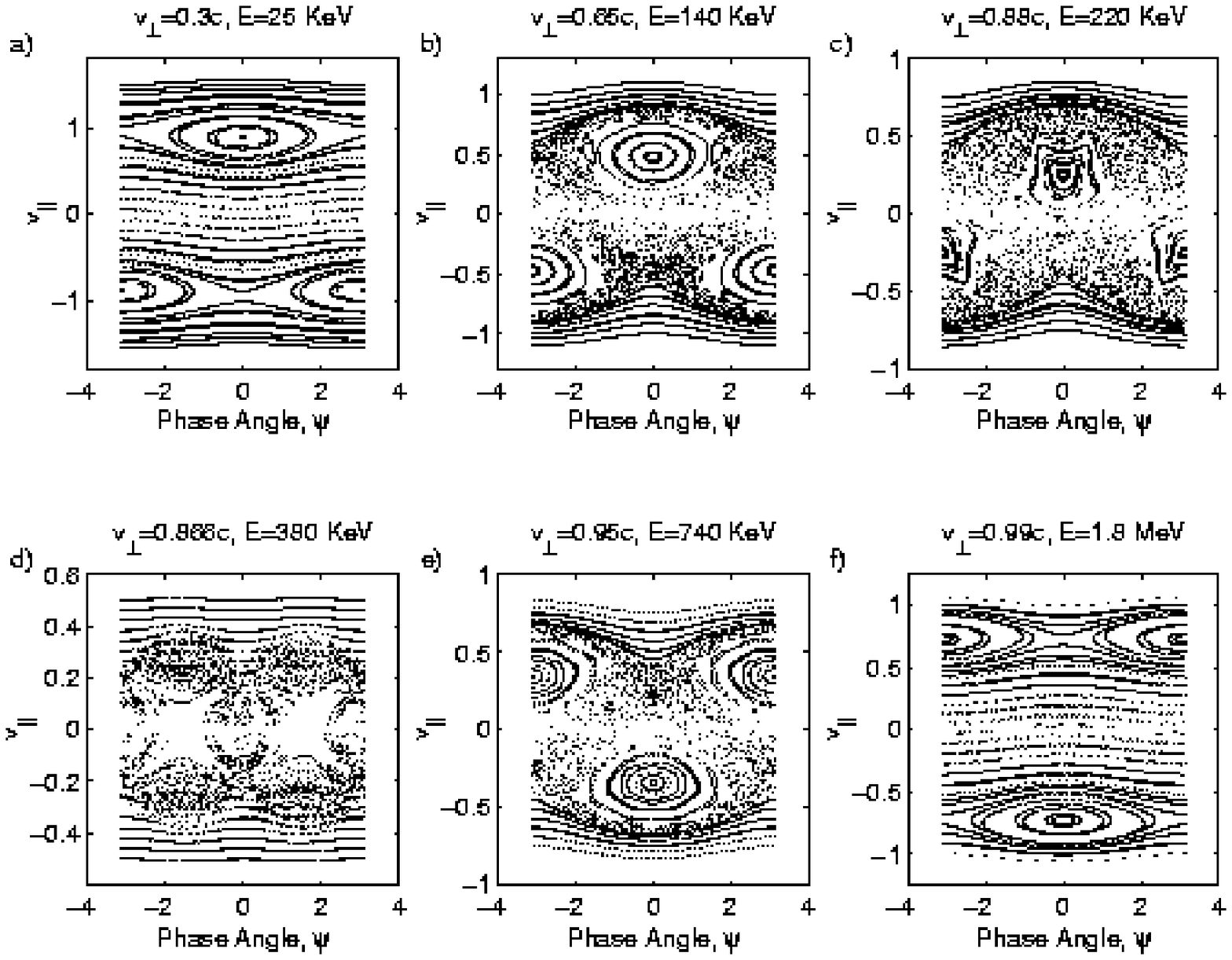}
\end{figure}

\begin{figure}[htbp]
\centering
\includegraphics[width=1.0\textwidth]{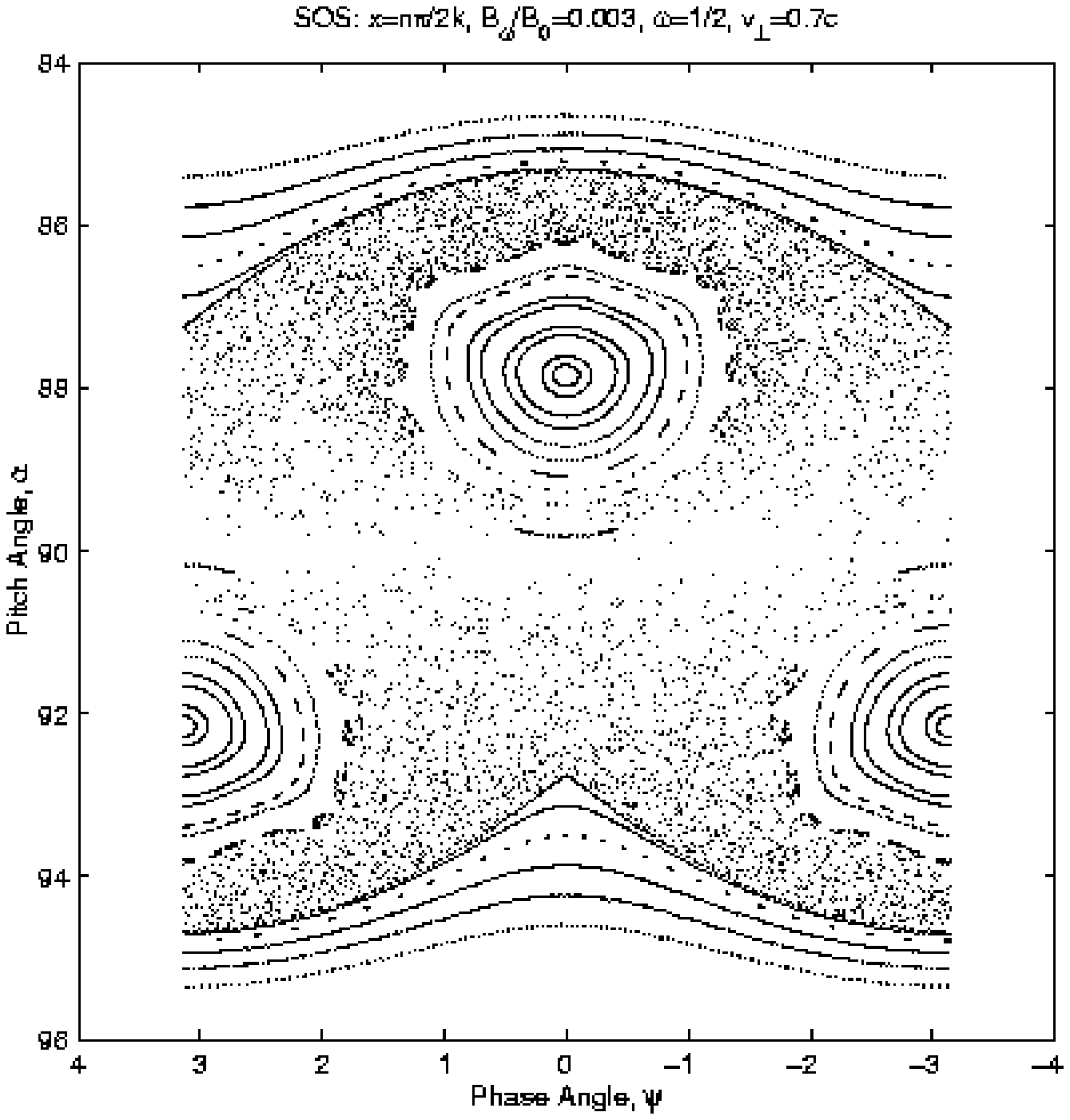}
\end{figure}

\begin{figure}[htbp]
\centering
\includegraphics[width=1.0\textwidth]{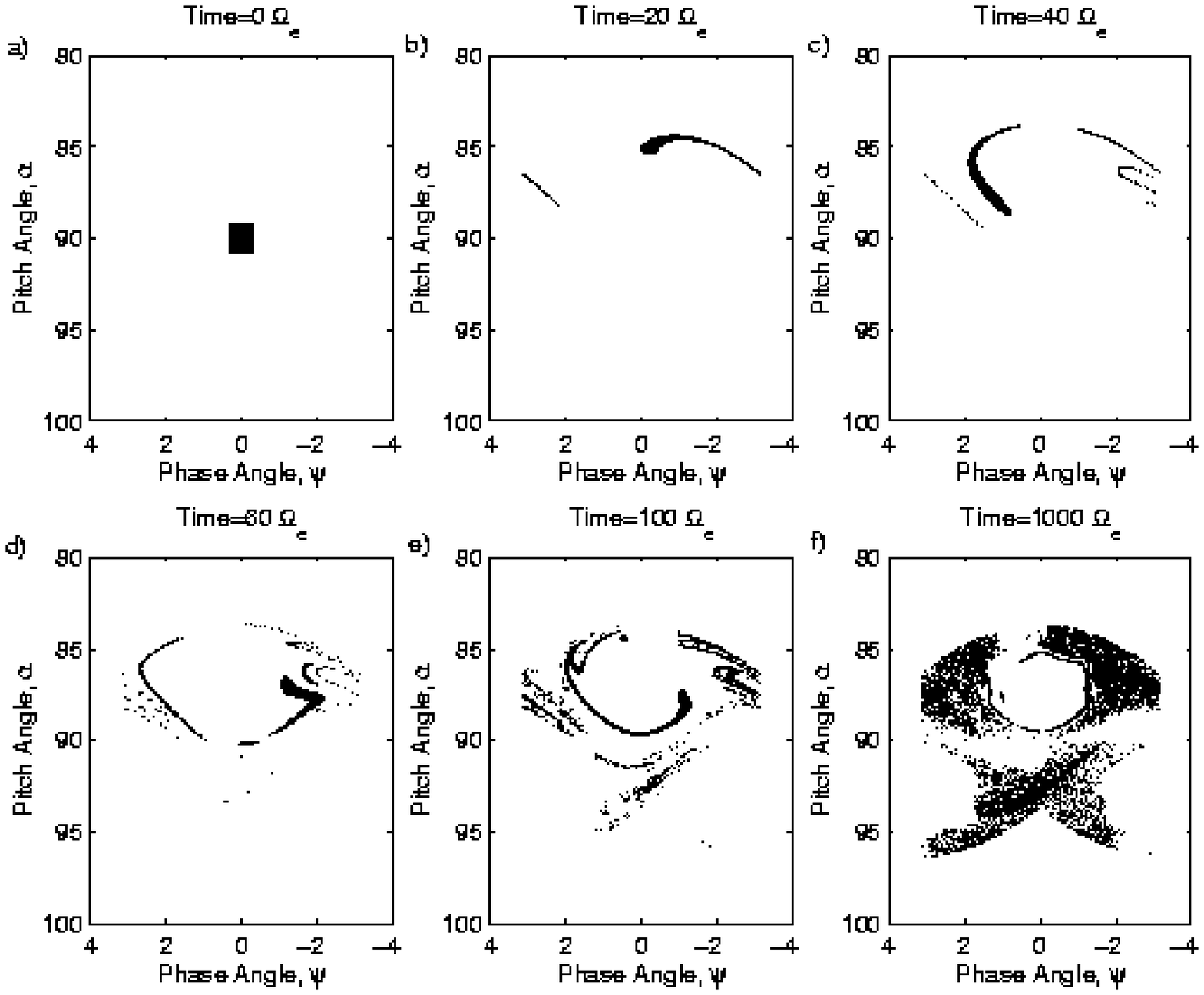}
\end{figure}

\begin{figure}[htbp]
\centering
\includegraphics[width=1.0\textwidth]{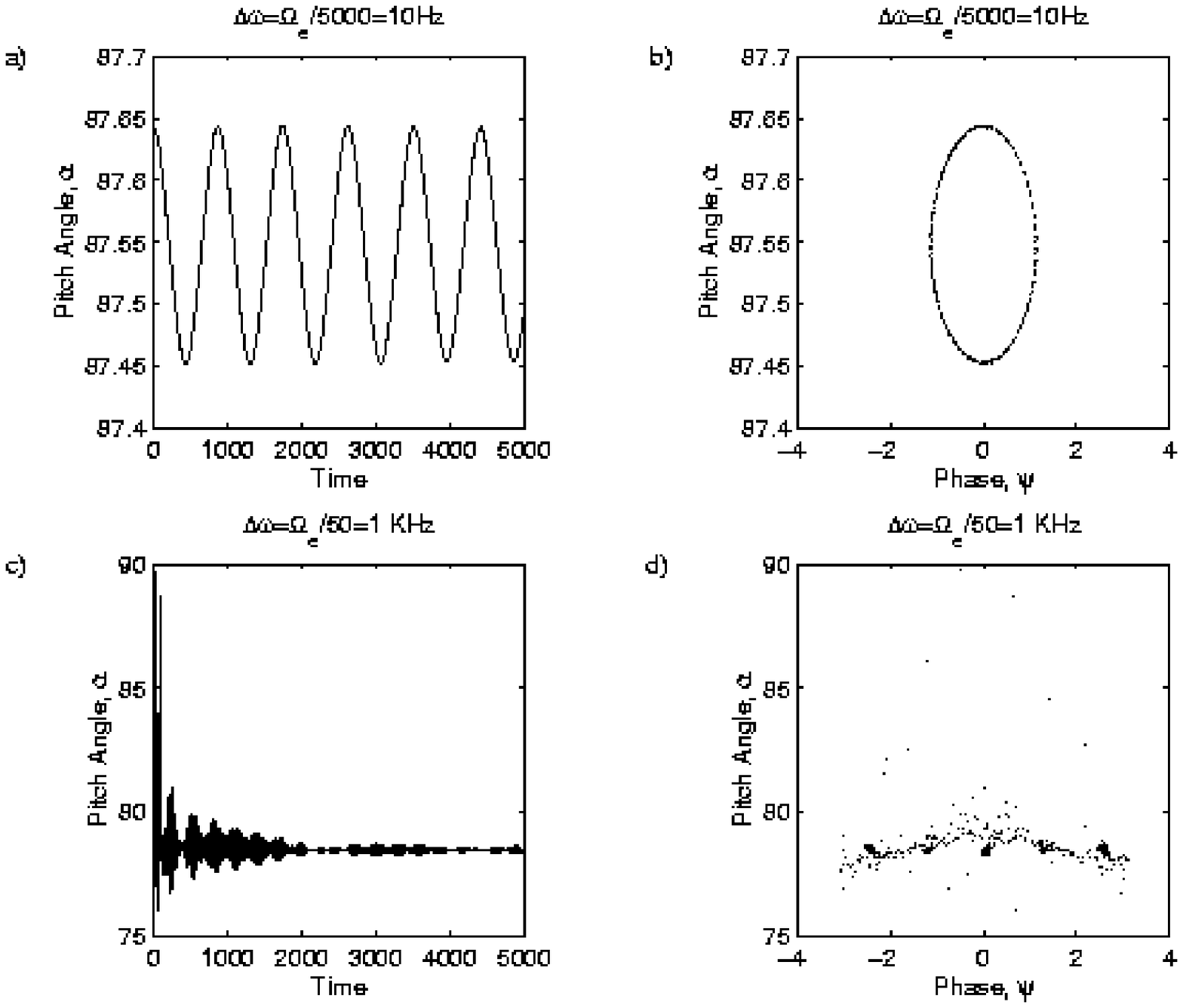}
\end{figure}

\begin{figure}[htbp]
\centering
\includegraphics[width=1.0\textwidth]{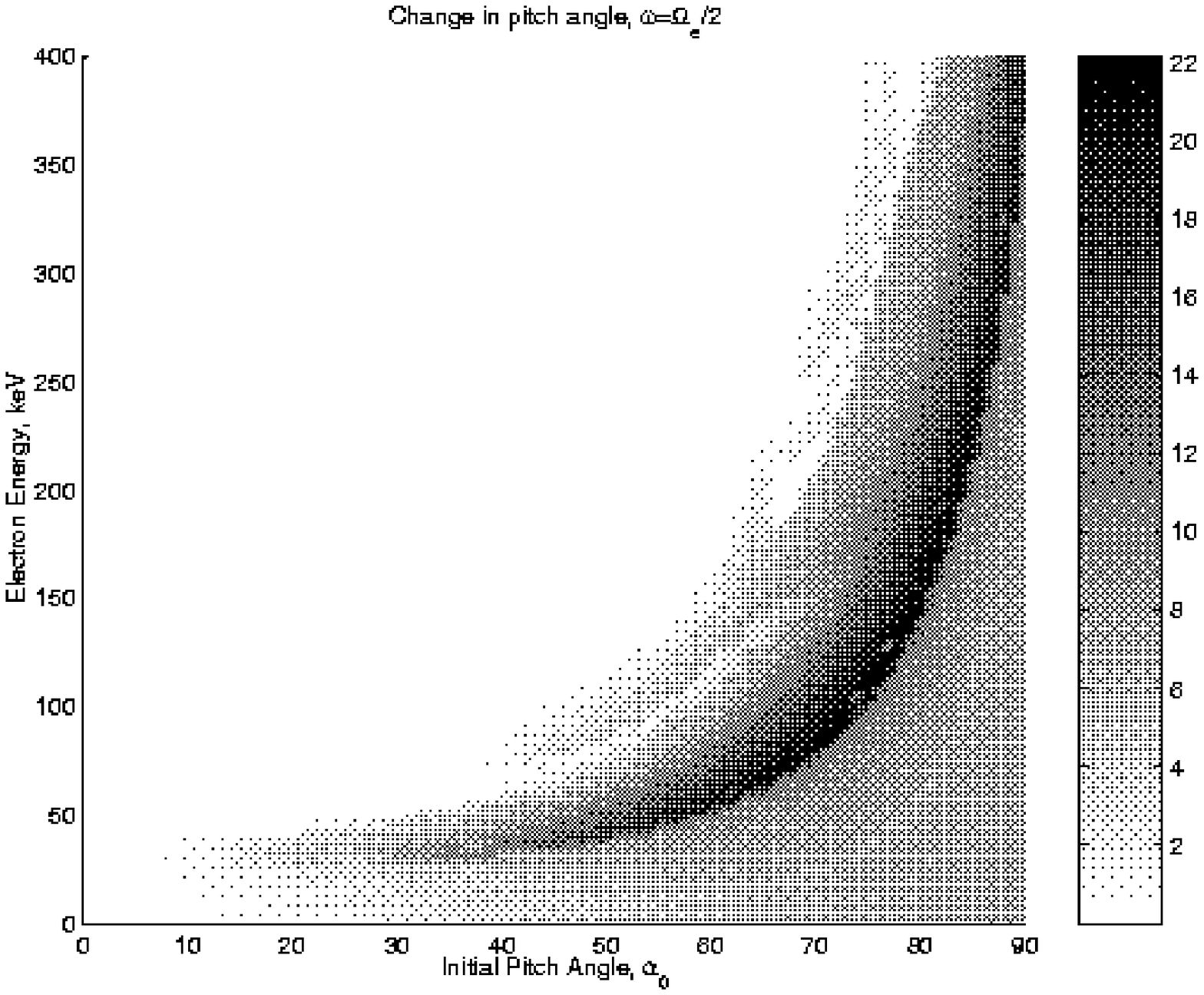}
\end{figure}

\begin{figure}[htbp]
\centering
\includegraphics[width=1.0\textwidth]{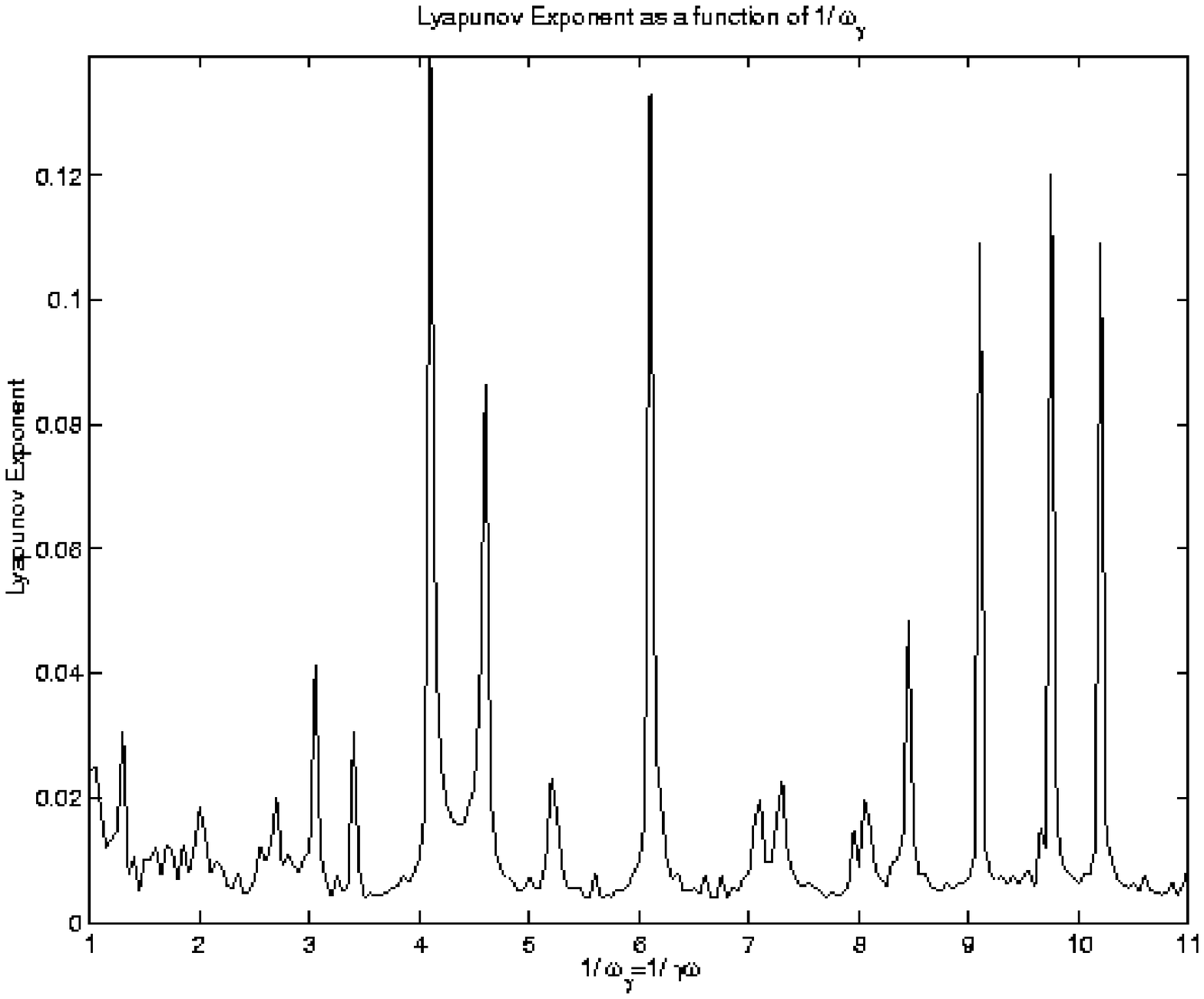}
\end{figure}

\end{document}